\documentclass[conference]{IEEEtran}
\IEEEoverridecommandlockouts
\usepackage{cite}
\usepackage{amsmath,amssymb,amsfonts,bm,amsthm}
\usepackage{algorithm}
\usepackage{algpseudocode}
\usepackage{graphicx}
\usepackage{textcomp}
\usepackage{xcolor}
\usepackage{diagbox}
\usepackage{multirow}
\newtheorem*{remark}{Remark}
\def\BibTeX{{\rm B\kern-.05em{\sc i\kern-.025em b}\kern-.08em
    T\kern-.1667em\lower.7ex\hbox{E}\kern-.125emX}}
\begin{document}

\title{A Multi-Head Ensemble Multi-Task Learning Approach for Dynamical Computation Offloading \\
}

\author{
\IEEEauthorblockN{Ruihuai Liang\IEEEauthorrefmark{1}, Bo Yang\IEEEauthorrefmark{1}, Zhiwen Yu\IEEEauthorrefmark{1},  Xuelin Cao\IEEEauthorrefmark{2}, Derrick Wing Kwan Ng\IEEEauthorrefmark{3}, and  Chau Yuen\IEEEauthorrefmark{4}} 
  
\IEEEauthorblockA{\IEEEauthorrefmark{1}School of Computer Science, Northwestern Polytechnical University, Xi'an, Shaanxi, 710129, China} 
\IEEEauthorblockA{\IEEEauthorrefmark{2}School of Cyber Engineering, Xidian University, Xi'an, Shaanxi, 710071, China}
\IEEEauthorblockA{\IEEEauthorrefmark{3}School of Electrical Engineering and Telecommunications, University of New South Wales, Sydney, Australia} 
\IEEEauthorblockA{\IEEEauthorrefmark{4}School of Electrical and Electronics Engineering, Nanyang Technological University, Singapore}

}

\maketitle

\begin{abstract}
Computation offloading has become a popular solution to support computationally intensive and latency-sensitive applications by transferring computing tasks to mobile edge servers (MESs) for execution, which is known as mobile/multi-access edge computing (MEC). To improve the MEC performance, it is required to design an optimal offloading strategy that includes offloading decision (i.e., whether offloading or not) and computational resource allocation of MEC. The design can be formulated as a mixed-integer nonlinear programming (MINLP) problem, which is generally NP-hard and its effective solution can be obtained by performing online inference through a well-trained deep neural network (DNN) model. However, when the system environments change dynamically, the DNN model may lose efficacy due to the drift of input parameters, thereby decreasing the generalization ability of the DNN model. To address this unique challenge, in this paper, we propose a multi-head ensemble multi-task learning (MEMTL) approach with a shared backbone and multiple prediction heads (PHs). Specifically, the shared backbone will be invariant during the PHs training and the inferred results will be ensembled, thereby significantly reducing the required training overhead and improving the inference performance. As a result, the joint optimization problem for offloading decision and resource allocation can be efficiently solved even in a time-varying wireless environment. Experimental results show that the proposed MEMTL outperforms benchmark methods in both the inference accuracy and mean square error without requiring additional training data\footnote{This paper will be presented at the IEEE Globecom conference 2023.}. 
\end{abstract}


\section{Introduction}\label{sec_introduction}
As a new computing paradigm, mobile/multi-access edge computing (MEC) provides cloud computing services at the edge of the network close to mobile terminals (MTs), avoiding the drawbacks such as the long propagation latency introduced by conventional mobile cloud computing~\cite{base_survey, CO_dl_survey, CO_MEC_survey_2}. Indeed, the popularity of MEC has promoted the rapid growth of mobile applications, in which computation offloading plays a critical role. 
To achieve an optimal balance between execution time and resource consumption, and to improve the experience of MTs, computation offloading shifts computationally intensive or latency-sensitive tasks to a MEC server (MES) or cloud server~\cite{CO_MEC_survey_1}. Considering optimal designs such as binary offloading decision, resource allocation, mobility management, content caching, security and privacy, computation offloading optimization can be formulated as a mixed-integer nonlinear programming (MINLP) problem, which is generally NP-hard~\cite{NP_ref} and thus difficult to solve efficiently. 

Existing computation offloading approaches can be roughly divided into two categories~\cite{CO_survey_taxonomy}. The first category is the conventional numerical optimization~\cite{CO_convex_opt_example} approaches that obtain solutions through repeated iterations. These methods usually require preset parameters and introduce excessive computational complexity, even though they may usually obtain only sub-optimal solutions. Furthermore, once the network environment alters, the same iterative procedure needs to be carried out again to update the solution. The second category is the artificial intelligence (AI)-based approaches~\cite{base_work,CO_drl_1}. This kind of approaches can learn the potential knowledge of the task and can directly infer the optimal solutions in a near real-time manner but with low complexity. However, it is worth noting that there still exists some potential challenges. For instance, the AI model training process often requires a large amount of labeled training data for offline learning and the online inference performance is often susceptible. 

Specifically, a multi-task learning feedforward neural network (MTFNN) was proposed in~\cite{base_work} to jointly optimize the offloading decision and computational resource allocation, which has offered remarkable results. However, the MTFNN model was offline trained under the training dataset collected in a specific network condition, thereby suffering from the issue of insufficient generalization. It is widely known that with more training dataset available,  better inference performance can be usually achieved. Due to the relatively high computational complexity of exhaustive searching method that has been exploited for training dataset generation, the size of training dataset is difficult to expand effectively and efficiently. Furthermore, even if a large amount of training dataset can be generated, static model structure still has poor scalability in dealing with dynamical network environments.

Motivated by drawbacks of MTFNN, we present a multi-head ensemble multi-task learning (MEMTL) approach for the joint optimization of offloading decision and resource allocation in a dynamical network environments. By establishing multiple prediction heads (PHs) with a shared backbone network, MEMTL not only outperforms the baseline in terms of accuracy and mean square error (MSE), but also has better scalability to cope with dynamic computation offloading. Moreover, the proposed MEMTL does not require additional training data, while only having a negligible increase in model storage space and inference time. 


\section{System Model}\label{section_2}
\subsection{Overview}\label{system_model_overview}
\begin{figure}[h]
\centering
\includegraphics[width=2.570in, height=1.354in]{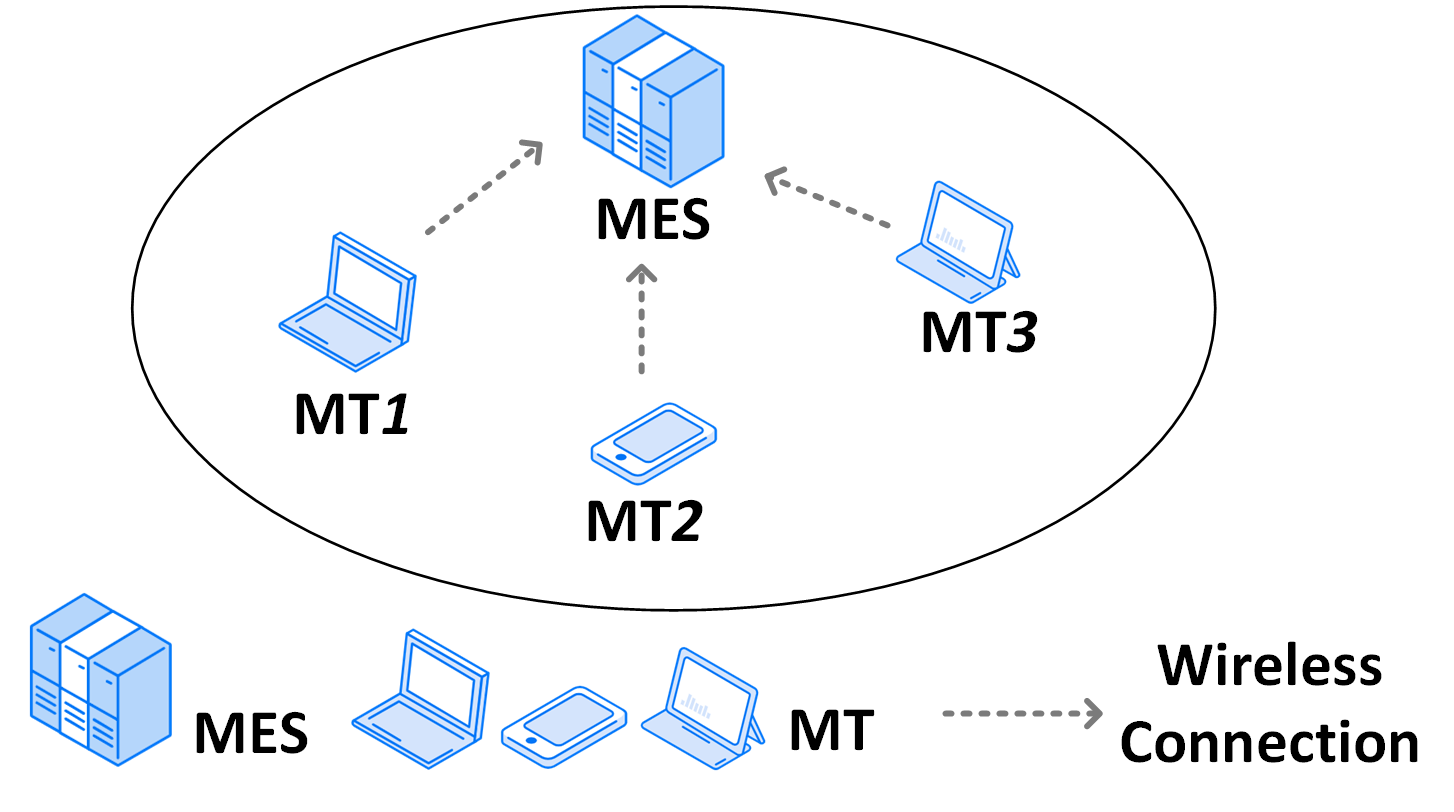}
\caption{An example of a multi-user single-server MEC system with one MES and three MTs.}
\label{fig1_mec}
\end{figure}

Considering a multi-user single-server MEC system that consists of one MES and $N$ MTs, i.e., ${\cal N}=\{MT_1, MT_2, ..., MT_N\}$. Each MT is wirelessly connected with the MES and different offloading connections could introduce various amount of transmission overheads due to the diversity of the transmission power and the distances, as illustrated in Fig.~\ref{fig1_mec}. Provided that the orthogonal multiple access~\cite{NOMA_survey} is employed as the multiple access scheme for the wireless connections, so the whole wireless bandwidth is divided into $N$ equal sub-bands, each of which serves transmission without interferences. Specifically, we define the upload transmission rate and the download transmission rate of the connection between the MES and $MT_n$ as $u_n$ and $d_n$, respectively. In the considered MEC scenario, the optimal offloading strategy includes the offloading decision (i.e., whether to offload or process locally) and the computational resource allocation of the MES. 

Specifically, regarding the offloading method, we focus on the binary offloading strategy~\cite{base_survey}, which indicates that the job is either processed locally or totally offloaded to the MES for further processing. We assume that all MTs make offloading decisions at each discrete time simultaneously and each MT considers only one job to be offloaded at a time, which is indivisible. The two options for offloading are formulated as follows:
\begin{equation}\label{e_decision_option}
    D_{n}=\left\{\begin{matrix}
        1 & {\rm if }\ MT_n \ {\rm offloads}, \\ 
        0 & \rm otherwise.
        \end{matrix}\right. \ \forall n \in [1,N].
\end{equation}

To characterize the offloading strategy of each MT, we define an $N$-dimensional offloading decision vector, i.e., 
\begin{equation}\label{e_decision_vec}
    \mathbf{D}=\{D_{1}, ...,D_{n},..., D_{N}\}, \ \forall n \in \cal N.
\end{equation}

In addition, we define another $N$-dimensional vector to indicate the computational resource allocation of the MES, 
\begin{equation}\label{e_allocation_vec}
    \mathbf{R}=\{R_{1}, ...,R_{n},..., R_{N}\}, \ \forall n \in \cal N.
\end{equation}
where $R_{n} \in [0,1]$ denotes the proportion of computational resource allocated by the MES to $MT_n$, such that $\sum_{n=1}^{N}R_{n}\leq 1$ holds. 


\subsection{Problem Formulation}
The objective of computation offloading is to minimize the overall cost that includes the execution latency and energy consumption. Provided that $c_n(c_n\ge 0)$ (clock cycle) denotes the required computational resource and $r_{n}^{local}$ (clock cycle) refers to the available computational resource of $MT_n$, the cost of local execution of $MT_n$ is calculated as 
\begin{equation}\label{e_cost_local}
{\cal C} _{\rm local}^{n}=(1-\alpha)\kappa \left ( r_{n}^{local} \right )^2 c_n + \alpha \frac{c_n}{r_{n}^{local}}, \ \forall n \in {\cal N},
\end{equation}
where ${\cal C}_{\rm local}^{n}$ denotes the overall cost of locally processing on $MT_n$, $\kappa$ denotes the energy efficiency parameter that mainly depends on the hardware chip architecture~\cite{kappa}, and $\alpha\in [0,1]$ determines the emphasis on computational delay and power consumption. For example, the delay cost is more important when $\alpha$ is close to 1 and the power cost is more important when $\alpha$ is close to 0. 

Assume that $p_n$ denotes the size of data that requires for uploading, $q_n$ indicates the size of the results that are returned after processing, so the cost of offloading is 
\begin{equation}\label{e_cost_offload}
\begin{split}
    {\cal C}_{\rm offload}^{n}=&(1-\alpha)\left(\frac{P_{u}^{n} p_n}{u_{n}}+\frac{P_{e}^{n} c_n}{r_{n}^{offload}}+\frac{P_{d}^{n} q_n}{d_{n}}\right) \\ 
    &+\alpha \left(\frac{p_n}{u_n}+\frac{c_n}{r_{n}^{offload}}+\frac{q_n}{d_n}\right),\ \ \forall n \in {\cal N},
\end{split}
\end{equation}
where $P_{u}^{n}$, $P_{e}^{n}$, $P_{d}^{n}$ represent the power consumption of job uploading, execution and downloading, respectively, $r_{n}^{offload}$ (clock cycle) denotes the computational resource allocated from MES to $MT_n$. 

Combining (\ref{e_cost_local}) and (\ref{e_cost_offload}), the weighted sum cost can be calculated as 
\begin{equation}\label{e_overall_cost}
   {\cal C}_{\rm total}=\sum_{n\in{\cal N}} \left ({1-D_{n}}\right ) {\cal C}_{\rm local}^{n}+ D_{n}  {\cal C}_{\rm offload}^{n},
\end{equation}
where ${\cal C}_{\rm local}^{n}$ denotes the total cost of $MT_n$ processing the job locally and ${\cal C}_{\rm offload}^{n}$ indicates the total cost of offloading. Consequently, the optimization problem is formulated as 
\begin{subequations}\label{e_obj_op_func}
\begin{align}
    &  \mathbb{P}: \underset{\{\mathbf{D},\mathbf{R}\}}{{\rm min}}\ {\cal C}_{\rm total} \notag \\
    &s.t.\ \mathbf{C1}: \ D_{n}\in \{0,1\}, \ \forall n \in {\cal N}, \\
    &\ \ \ \ \ \mathbf{C2}: \ (1-D_{n})\frac{c_n}{r_{n}} + D_{n}\left(\frac{p_n}{u_n}+\frac{c_n}{r_{n}^{offload}}+\frac{q_n}{d_n}\right)\leq \theta_n, \\
    &\ \ \ \ \ \mathbf{C3}: \ R_{n} \in [0,1], \\
    &\ \ \ \ \ \mathbf{C4}: \ \sum_{i=1}^{N}R_{i}\leq 1.
\end{align}
\end{subequations}
The constraint $\mathbf{C1}$ denotes that the offloading decision only has two options, i.e., $D_{n}\!=\!0$ or $D_{n}\!=\!1$. $\mathbf{C2}$ indicates that the maximum tolerable delay of the job of $MT_n$ is $\theta_n$, which should be satisfied. $\mathbf{C3}$ means that the resource allocation proportion should be in the range of 0 and 1, and $\mathbf{C4}$ indicates the sum of allocation proportions should not be larger than 1. 

\begin{figure*}[ht]
\centering
\includegraphics[width=6.00in, height=1.63in]{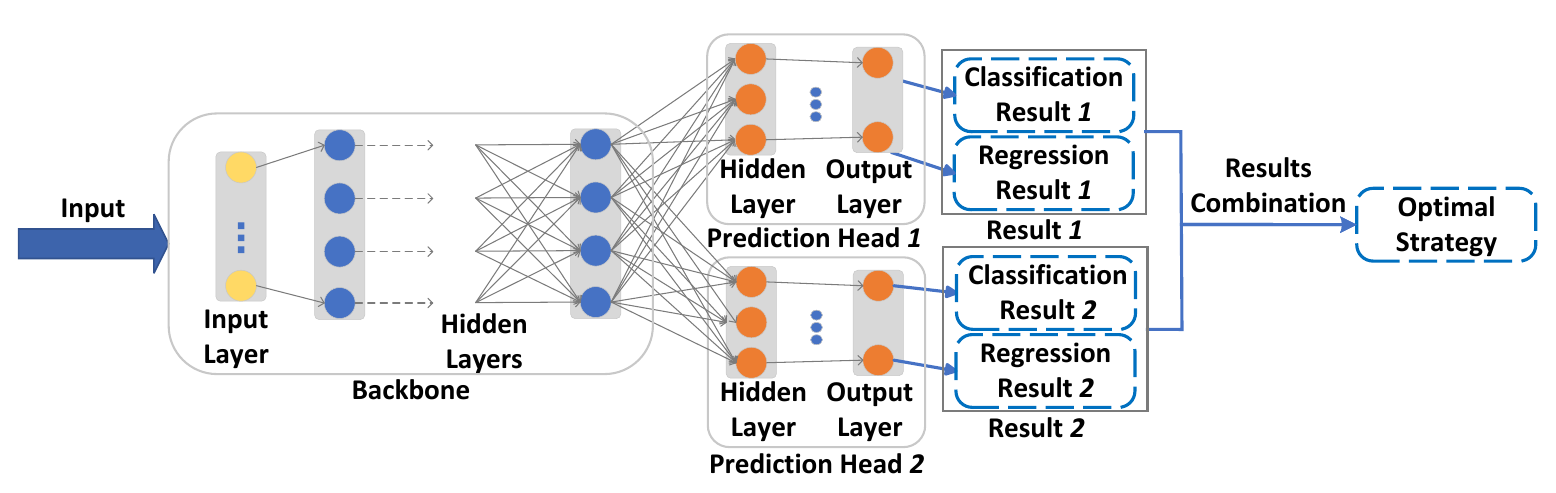}
\caption{The structure of the proposed MEMTL model with $2$ PHs.}
\label{fig2_memtl}
\end{figure*} 

We can observe that the formulated multi-dimensional multi-objective problem is actually a mixed-integer nonlinear programming (MINLP) problem, which is generally NP-hard and can be compartmentalized through the Tammer~\cite{Tammer} decomposition method. Specifically, the optimal solution of the joint optimization problem can be obtained as $\mathbf{S^*}\!=\!\{\mathbf{D^*}, \mathbf{R^*}\}$, where $\mathbf{D^*}$ indicates the optimal offloading decision vector and $\mathbf{R^*}$ indicates the optimal computational resource allocation ratio vector. Considering that $\mathbf{D}$ is a binary vector and $\mathbf{R}$ has decimal features in [0, 1], we can design multi-task learning that consists of a classification task and a regression task. 

After formulating the original problem as an hierarchical optimization problem of minimizing the objective function, we can find that it can be further simplified as a problem to minimize a function $f(x,y)$ with a given $x$ and an output $y$, where $x$ refers to the features of all MT jobs and network environment parameters, while $y$ refers to the optimal solution. Although performing offline training based on a large amount of samples via a single neural network model can achieve good accuracy on the test set, this approach suffers from a severe performance degradation in dynamical network environments. In contrast, by training multiple component models independently and combining them for inference, ensemble methods reach better performance than the single network counterpart~\cite{m_head_why, ensemble_base_review}. In next section, we provide the detailed design of the proposed model. 

\section{Multi-Head Ensemble Multi-Task Learning}
In this section, we firstly present the proposed multi-head ensemble multi-task learning (MEMTL) approach in detail, followed by some analytical results for quantifying the potential performance gain. 

\subsection{Model Design}
The structure of the proposed MEMTL model is shown in Fig.~\ref{fig2_memtl}, where two PHs are considered as an example. Specifically, the input to the model includes the vector of the jobs and environment parameters, the body structure of the MEMTL comprises multiple dense hidden layers, and the output layer consists of two tasks: classification task and regression task, which aims at inferring $\mathbf{D}^*$ and $\mathbf{R}^*$, respectively. Ensemble improves model scalability, but excessive computation complexity will be introduced by building a group of complete models. This hinders its practical implementation in dynamical offloading environment with limited computational and storage resources. 

Instead of training multiple MTFNN models under different cases, we find that there exist common features that can be extracted from the knowledge learned by each MTFNN model and these common features can be considered as a common invariant part (i.e., the backbone in Fig.~\ref{fig2_memtl}) such that only small-scale expansion on ensemble components needs to be undertaken. Inspired by~\cite{m_head_search, m_head_why}, we employ the multi-head neural network structure which has multiple PHs with a small number of tunable parameters and a shared backbone with ``frozen'' parameters, as illustrated in Fig.~\ref{fig2_memtl}. Specifically, the backbone is a continuous series of multiple layers starting from the input layer. During the components training process, the constructed backbone's parameters are invariant and thus do not participate in the gradient descent of PHs, which reduces both the computational and time consumption. In particular, each PH is composed of a few network layers and the output layer, which lay after the backbone. For ease of understanding, each combination of a PH and the backbone can be regarded as an independent MTFNN. The backbone is exploited to extract common latent features from the input, which will be considered as the input to the PHs. Also, the outputs from the PHs are final inferring results based on the information extracted by the backbone. It is worthy noting that only PHs participate in the gradient descent process of components training, which can help to significantly reduce the training cost that facilitates the implementation of online training.

\subsection{Training and Prediction}\label{sec_memtl_train_pre}
After introducing the model structure, in this subsection we present the offline training and online prediction procedures. 

In \textbf{Algorithm~\ref{al_memtl_offline_train}}, we illustrate the offline training process of the MEMTL model with $M$ PHs that share a backbone (denoted as $\cal B$), where the loss function consists of MSE for the regression task and cross entropy loss for the classification task. To ensure the ability to extract the common features properly and accurately, the backbone is trained based on the whole dataset. For the training of PHs, we employ the bootstrap sampling which performs repeated replacement sampling on the original dataset, and generate sub-datasets with the same size. By creating multiple datasets through bootstrap sampling, each PH is trained based on a slightly different subset of the original dataset, thereby helping to mitigate overfitting and improve generalization performance.
\renewcommand{\algorithmicrequire}{\textbf{Input:}}  
\renewcommand{\algorithmicensure}{\textbf{Output:}}
\begin{algorithm} 
\caption{MEMTL Offline Training}        
\small
\label{al_memtl_offline_train}
\begin{algorithmic}[1]
\Require Training dataset $\cal D$ that contains $T$ labelled data samples, the number of the PHs $M$;

\Ensure Trained MEMTL model with a common backbone $\cal B$ and $M$ PHs ${\cal H} = \{h_1, ..., h_M\}$;

\State Perform sampling from $\cal D$ by bootstrapping and generate $M$ equal-sized datasets $\{D_1, ..., D_M\}$, where $D_m,\ \forall m \in [1,M]$, has $T$ samples;
\State Set the backbone $\cal B$ and the PH $h_1$ to $trainable=True$, train the $B$ and $h_1$ on the source dataset $\cal D$;
\State Store the trained backbone $\cal B$ and reinitialize $h_1$, and set $\cal B$ to $trainable=False$;
\For{$m=1$ to $M$}
\State Set $h_m$ to $trainable=True$ and other PHs to $trainable=False$;
\State Train $h_m$ on the dataset $D_m$;
\EndFor
\State Return the trained MEMTL model with backbone $\cal B$ and PHs $\cal H$.
\end{algorithmic}
\end{algorithm}

\vspace{-1.75mm}
\renewcommand{\algorithmicrequire}{\textbf{Input:}}  
\renewcommand{\algorithmicensure}{\textbf{Output:}}
\begin{algorithm}
\caption{MEMTL Online Prediction}        
\small
\label{al_memtl_online_pre}
\begin{algorithmic}[1]
\Require Input parameter vector $x$, the trained MEMTL model $\cal H$;

\Ensure The inferred optimal offloading strategy $\textbf{S}^*$;

\State Input $\mathbf{x}$ to $\cal H$, obtain the output of each PH, i.e., $\{\textbf{S}_1, ..., \textbf{S}_M\}$ ;
\State Select the optimal offloading strategy $\mathbf{S}^*$ that reaches the minimum overall cost, i.e.,  $\textbf{S}^*=\underset{\textbf{S}_i,i\in [1, M]}{\rm argmin} \ {\cal C}_{\rm total}$.

\end{algorithmic}
\end{algorithm} 
\vspace{-0.8mm}

In \textbf{Algorithm~\ref{al_memtl_online_pre}}, we present the online prediction method to infer the optimal offloading strategy. Specifically, multiple component PHs conduct online inference based on the input and generate their respective output strategies, respectively. Then, we choose the optimal offloading strategy from the multiple PHs that minimizes the overall cost. 

\subsection{Theoretical Analysis}\label{sec_theo_analysis}
To demonstrate the performance enhanced by ensemble learning,  we present the theoretical analysis on the average generalization error of the proposed MEMTL method by employing the error-ambiguity decomposition method~\cite{error_ambiguity_decomposition}. 

Suppose that a MEMTL model $\cal H$ with $M$ PHs $\{h_1, ..., h_M\}$ learns the target function $f{:} \ \mathbf{R}^d\rightarrow \mathbf{R}^g$, where $d$ is the input dimension and $g$ is the output dimension. During the inference, for the sample $\mathbf{x}$ with the feature distribution $p(\mathbf{x})$, each PH returns its inferring result $h_i(\mathbf{x})$. Specifically, the prediction result given by $\cal H$ is denoted as $\cal H(\mathbf{x})$, we define the MSE on the single sample $\mathbf{x}$ of $\cal H$ and $h_i$ as $\xi({\cal H}|\mathbf{x})$ and $\xi(h_i|\mathbf{x})$, respectively, then we can obtain
\begin{subequations}
\begin{align}
    \xi({\cal H}|\mathbf{x})\overset{\underset{\mathrm{def}}{}}{=}&\frac{1}{g}(f(\mathbf{x})-{\cal H}(\mathbf{x}))^2, \label{e_raw_err_H}\\
    \xi(h_i|\mathbf{x})\overset{\underset{\mathrm{def}}{}}{=}&\frac{1}{g}(f(\mathbf{x})-h_i(\mathbf{x}))^2, \ i\in [1,M], \label{e_raw_err_h}
\end{align}
\end{subequations}
where the error refers to the discrepancy between the inference output and the ground truth. 

The ambiguity is defined as the variation of the outputs of ensemble members over the unlabelled data, i.e., it quantifies the disagreement among the components. We define the ambiguity between $h_i$ and $h_o$ as $\chi (h_i)$ 
\begin{equation}\label{e_ambiguity_define}
    \chi (h_i)\overset{\underset{\mathrm{def}}{}}{=}\xi(h_i|\mathbf{x})-\xi(h_o|\mathbf{x}), \ \forall i,o\in [1,M]
\end{equation}
where $h_o$ denotes the optimal PH with the smallest MSE such that $\xi(h_i|\mathbf{x})\ge \xi(h_o|\mathbf{x})$ always holds and the $\chi(h_i)$ is non-negative. 

Over the entire feature distribution domain $p(\mathbf{x})$, the generalization error of $\cal H$ and $h_i$ are defined as 
\begin{subequations}
\begin{align}
    \zeta({\cal H})&=\int \xi ({\cal H}|\mathbf{x}) p(\mathbf{x})d\mathbf{x}, \label{e_gene_err_H} \\
    \zeta(h_i)&=\int \xi (h_i|\mathbf{x}) p(\mathbf{x})d\mathbf{x},\ \forall i \in [1,M]. \label{e_gene_err_hi}
\end{align}
\end{subequations}

The \textbf{Algorithm~\ref{al_memtl_online_pre}} chooses the output with the minimum cost as the final result, which is equivalent to choosing the output with the smallest MSE. As the optimal PH with the smallest MSE is $h_o$, the $\xi({\cal H}|\mathbf{x})$ can be re-expressed as 
\begin{equation}\label{e_err_H_re_exp}
\begin{split}
    \xi({\cal H}|\mathbf{x})=\xi({h_o}|\mathbf{x})&=\xi({h_i}|\mathbf{x})-[\xi({h_i}|\mathbf{x})-\xi({h_o}|\mathbf{x})], \\
    &\ \ \ \ \ \ \ \ \ \ \ \ \ \ \ \ \ \ \ \ \ \ \ \forall i,o\in [1,M].
\end{split}
\end{equation}

Substituting (\ref{e_err_H_re_exp}) into (\ref{e_gene_err_H}), $ \zeta({\cal H})$ can be deduced as 
\begin{equation}\label{e_gene_err_H_mid_1}
\begin{split}
    \zeta({\cal H})=\int [\xi({h_i}|\mathbf{x})-(\xi({h_i}|\mathbf{x})-\xi({h_o}|\mathbf{x}))]p(\mathbf{x})d\mathbf{x},\\
    \ \ \ \ \ \ \ \ \ \ \ \ \ \ \ \ \ \ \ \ \ \forall i,o\in [1,M].
\end{split}
\end{equation}

Multiplying both sides of equation (\ref{e_gene_err_H_mid_1}) by $M$ at the same time, and specializing the $M$ $i$'s on the right into $i=\{1,...,M\}$, we obtain 
\begin{equation}\label{e_gene_err_H_mid_11}
\begin{split}
    &M*\zeta({\cal H})=\int [\xi({h_1}|\mathbf{x})-(\xi({h_1}|\mathbf{x})-\xi({h_o}|\mathbf{x}))]p(\mathbf{x})d\mathbf{x} + ...\\
    &\ \ \ \ \ \ \ \ \ +\int [\xi({h_M}|\mathbf{x})-(\xi({h_M}|\mathbf{x})-\xi({h_o}|\mathbf{x}))]p(\mathbf{x})d\mathbf{x}\\
    &=\sum_{i=1}^{M}\int\xi({h_i}|\mathbf{x})p(\mathbf{x})d\mathbf{x}-\sum_{i=1}^{M}\int(\xi({h_i}|\mathbf{x})-\xi({h_o}|\mathbf{x}))p(\mathbf{x})d\mathbf{x}.
\end{split}
\end{equation}

Furthermore, by dividing both sides of equation by $M$ (\ref{e_gene_err_H_mid_11}), it yields 
\begin{equation}\label{e_gene_err_H_mid_2}
\begin{split}
    \zeta({\cal H})&=\frac{1}{M}\sum_{i=1}^{M}\int\xi({h_i}|\mathbf{x})p(\mathbf{x})d\mathbf{x}\\
    &\ \ -\frac{1}{M}\sum_{i=1}^{M}\int(\xi({h_i}|\mathbf{x})-\xi({h_o}|\mathbf{x}))p(\mathbf{x})d\mathbf{x}.
\end{split}
\end{equation}

Substituting (\ref{e_ambiguity_define}) and (\ref{e_gene_err_hi}) into (\ref{e_gene_err_H_mid_2}), we have 
\begin{equation}\label{e_err_ambi_decomp_mid_1}
    \zeta({\cal H})=\frac{1}{M}\sum_{i=1}^{M}\zeta(h_i)-\frac{1}{M}\sum_{i=1}^{M}\chi (h_i).
\end{equation}

\vspace{0.75mm}
For simplicity, we denote  $\overline{\xi}(h)=\frac{1}{M}\sum_{i=1}^{M}\zeta(h_i)$ and $
    \overline{\chi}(h)=\frac{1}{M}\sum_{i=1}^{M}\chi (h_i)$ as the average generalization error and the average ambiguity of all the $M$ PHs over the corresponding optimal PH, respectively.  
    
    Finally, substituting $\overline{\xi}(h)$ and $\overline{\chi}(h)$ into (\ref{e_err_ambi_decomp_mid_1}), we decompose the generalization error of $\cal H$ over the entire feature distribution domain as
\begin{equation}\label{e_err_ambi_de_final}
    \xi({\cal H})=\overline{\xi}(h)-\overline{\chi}(h),
\end{equation}
where $\overline{\chi}(h)$ is non-negative such that the ensemble error will not be higher than the average generalization error of all PHs. 
\begin{remark}
The higher accuracy and ambiguity of component models, the lower ensemble error can be obtained. 
\end{remark}

\section{Numerical Results and Discussions}\label{sec_res_discussion}

\begin{table*}[hbtp]
\caption{Performance Comparison of MEMTL and MTFNN}
\begin{center}
\begin{tabular}{|c|c|c|c|c|c|c|c|c|}
\hline
\multirow{2}*{\textbf{Number of MTs}} & \multicolumn{4}{c|}{\textbf{MTFNN}} & \multicolumn{4}{c|}{\textbf{MEMTL}} \\
\cline{2-9}
& \textbf{MSE} & \textbf{Accuracy} & \textbf{Inference time} & \textbf{Model size} & \textbf{MSE} & \textbf{Accuracy} & \textbf{Inference time} & \textbf{Model size} \\
\hline
$N$=2 & 0.025 & 93.5\% & 0.036 ms & 33.7 KB & 0.022 & 95.0\% & 0.063 ms & 49.1 KB \\
\hline
$N$=3 & 0.046 & 84.8\% & 0.028 ms & 35.1 KB & 0.036 & 89.4\% & 0.073 ms & 59.8 KB \\
\hline
$N$=4 & 0.057 & 77.8\% & 0.034 ms & 37.6 KB & 0.042 & 82.6\% & 0.059 ms & 68.0 KB \\
\hline
$N$=5 & 0.065 & 70.7\% & 0.026 ms & 38.1 KB & 0.050 & 77.9\% & 0.047 ms & 76.4 KB \\
\hline
\end{tabular}
\label{table_memtl_res}
\end{center}
\end{table*}

For the labeled dataset generation, we perform exhaustive search on randomly sampled input parameters and return ground truth paired with the input. To demonstrate the performance of the proposed MEMTL in dynamical environment, the source dataset is divided into multiple disjoint subsets. Specifically, we divide the dataset based on the mean of the features square of each sample, such that the parameters of each sample in a subset are significantly different from any sample from other subsets. That is to say, each subset can be regarded as a new environment compared with other subsets. Without loss of generality, we create two subsets during the experiments, one as training set and the other with less data as testing set represents the new environment. 

With the dataset prepared, we perform the experiments repeatedly in terms of two dimensions of variables: the number of MTs (denoted as $N$) and the number of PHs (denoted as $M$). All the experiments are performed based on the system of 11th Gen Intel(R) Core(TM) i5-11300H @ 3.10GHz (×16) processor. In each repetition, the baseline MTFNN and MEMTL are trained with the same training set and are evaluated on the same testing set. The implementation code of our proposed MEMTL method is available at: https://github.com/qiyu3816/MTFNN-CO.

We compare the performance between the MTFNN and the proposed MEMTL with 3 PHs, as illustrated in Tabel~\ref{table_memtl_res}, where we evaluate the MSE, accuracy, inference time (per sample) and model size. We observe that in all cases with different number of MTs, the MEMTL outperforms in terms of MSE and accuracy on the total testing set. This demonstrates the notable performance of MEMTL in dynamical environment with new input parameters. For the time and storage consumption caused by PHs, adding a single PH increases inference time by less than 0.02 ms and storage cost by about 10 KB which are usually acceptable in a MES. However, both the MEMTL and MTFNN inevitably suffer from performance degradation when the number of MTs increases, which indicates the considered task became more challenging. 

Although the performance of the proposed MEMTL model can be improved when the number of PHs increases, additional inference time and computational consumption will still be introduced. In order to investigate the appropriate number of PHs, we define the efficiency function of MEMTL as 
\begin{equation}\label{e_ef}
    \psi=\frac{\Delta \mathrm{mse} + \Delta \mathrm{accuracy}}{t},
\end{equation}
where $\psi(ms^{-1})$ denotes the improvement efficiency, $t(ms)$ denotes the inference time per sample. Compared to the MTFNN, $\Delta \mathrm{mse}$ and $\Delta \mathrm{accuracy}$ indicates the performance improvement of MEMTL in terms of MSE and accuracy, respectively. For instance, in the case with 2 MTs, we can observe from Table \ref{table_memtl_res} that $\Delta \mathrm{mse}=0.003$, $\Delta \mathrm{accuracy}=0.015$, $t=0.063$ ms. 

\begin{table}[hbtp]
\caption{Efficiency Function Results}
\label{table2_memtl_ef}
\centering
\begin{tabular}{|c|c|c|c|c|c|}
\hline
\diagbox{$N$}{$\psi$}{$M$} & $M=2$ & $M=3$ & $M=4$ & $M=5$ & $M=6$\\
\hline
$N=2$ & 0.2255 & 0.3047 & $\mathbf{0.4367}$ & 0.3884 & 0.2631\\
\hline
$N=3$ & 0.6518 & 0.7729 & 1.0722 & $\mathbf{1.1411}$ & 0.9573\\
\hline
$N=4$ & 1.0074 & 1.0556 & 1.1584 & $\mathbf{1.1905}$ & 1.0314\\
\hline
$N=5$ & 1.4915 & $\mathbf{1.8189}$ & 1.7160 & 1.5393 & 1.4549\\
\hline
\end{tabular}
\end{table}

In Table.~\ref{table2_memtl_ef}, we evaluate the efficiency $\psi$ of MEMTL, where $M$ denotes the number of PHs and $N$ indicates the number of MTs. As the bold best efficiencies show, to achieve the best efficiency, the number of PHs should be neither too many nor too little and generally the optimal PHs number in the table can be recommended. Additionally, as the number of MTs increases, we can observe that the efficiency of the MEMTL also increases. The reason behind this phenomenon is that the task that the model needs to learn is simpler when the number of MTs is smaller, so the MTFNN can already reach high accuracy that is larger than $90\%$. Therefore, the improvement provided by the MEMTL becomes limited. In this case, MEMTL provides more performance gain over MTFNN in more complex tasks. 

\begin{figure}[ht]
\centering
\includegraphics[width=3.6in, height=2.6in]{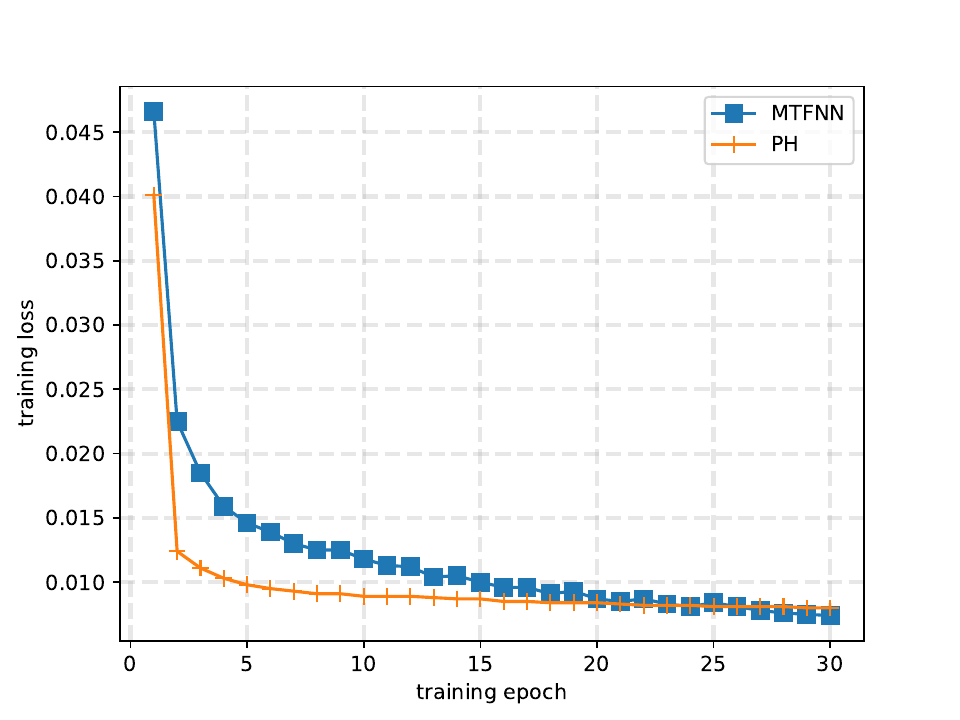}
\caption{Training loss evaluation between a single PH and a MTFNN from scratch. }
\label{fig_epoch_cost}
\end{figure}

In Fig.~\ref{fig_epoch_cost}, we compare the loss reduction processes in training MTFNN from scratch and training only PH. By setting the same learning rate as 0.001 and applying the same loss function, both the MTFNN model and the PH are trained for 30 epochs. We can observe that the loss of MEMTL was reduced to 0.01 in only 5 epochs, while the loss of MTFNN took at least 15 epochs, where the final losses are practically equal. From the above results, the fast convergence speed advantage of MEMTL online training PH is proved, and such excellent training efficiency also improves the potential to adapt to the dynamical network environment. 

\section{Conclusion and Future Work}
{ In this paper, we proposed a multi-head ensemble multi-task learning (MEMTL) approach for dynamical computation offloading in MEC environment. We formulated the joint optimization problem of offloading decision and computational resource allocation as a MINLP optimization problem, and then combine multi-head neural network and ensemble learning to design the MEMTL model and its training and inference strategies. By training the MEMTL model offline and performing an online inference test on input data which is significantly different from the training data, the test results show that the proposed MEMTL model significantly outperforms the baseline algorithm in terms of MSE and accuracy. By designing ensemble model for low-cost training and performing online inference, this not only offers inspiration for the model adaptation in dynamical environments in the fields such as communication and meteorology, but also provides a possible direction for large-scale model training and ensemble in distributed scenarios such as federated learning. As for our potential future works, we aim at addressing the problem of low efficiency of training data generation and the degradation of model performance for complex tasks with generative model and end-to-end trained models, respectively. }



\begin{thebibliography}{00}
\bibitem{base_survey} Y. Mao, C. You, J. Zhang, K. Huang and K. B. Letaief, ``A Survey on Mobile Edge Computing: The Communication Perspective," \textit{IEEE Commun. Surveys $\&$ Tutorials}, vol. 19, no. 4, pp. 2322-2358, Fourthquarter 2017.
\bibitem{CO_MEC_survey_2} K. Zhang, et al., ``Energy-Efficient Offloading for Mobile Edge Computing in 5G Heterogeneous Networks," \textit{IEEE Access}, vol. 4, pp. 5896-5907, 2016.
\bibitem{CO_dl_survey} X. Wang, Y. Han, V. C. M. Leung, D. Niyato, X. Yan and X. Chen, ``Convergence of Edge Computing and Deep Learning: A Comprehensive Survey," \textit{IEEE Commun. Surveys $\&$ Tutorials}, vol. 22, no. 2, pp. 869-904, Secondquarter 2020.
\bibitem{CO_MEC_survey_1} A. Islam, A. Debnath, M. Ghose, and S. Chakraborty, "A survey on task offloading in multi-access edge computing," \textit{Journal of Systems Architecture}, vol. 118, no. 102225, September 2021.
\bibitem{NP_ref} N.V. Sahinidis, ``Mixed-integer nonlinear programming 2018," \textit{Optimization and Engineering}, vol. 20, pp. 301-306, 2019.
\bibitem{CO_survey_taxonomy} H. Jin, M. A. Gregory and S. Li, "A Review of Intelligent Computation Offloading in Multiaccess Edge Computing," \textit{IEEE Access}, vol. 10, pp. 71481-71495, 2022.
\bibitem{CO_convex_opt_example} T. X. Tran and D. Pompili, ``Joint Task Offloading and Resource Allocation for Multi-Server Mobile-Edge Computing Networks," \textit{IEEE Transactions on Vehicular Technology}, vol. 68, no. 1, pp. 856-868, Jan. 2019.
\bibitem{base_work} B. Yang, X. Cao, J. Bassey, X. Li and L. Qian, ``Computation Offloading in Multi-Access Edge Computing: A Multi-Task Learning Approach," \textit{IEEE Transactions on Mobile Computing}, vol. 20, no. 9, pp. 2745-2762, Sept. 2021.
\bibitem{CO_drl_1} L. Huang, S. Bi and Y.-J. A. Zhang, ``Deep Reinforcement Learning for Online Computation Offloading in Wireless Powered Mobile-Edge Computing Networks," \textit{IEEE Transactions on Mobile Computing}, vol. 19, no. 11, pp. 2581-2593, Nov. 2020.
\bibitem{NOMA_survey}
L. Dai, B. Wang, Y. Yuan, S. Han, I. Chih-Lin, and Z. Wang, ``Non-orthogonal multiple access for 5G: solutions, challenges, opportunities, and future research trends,"\textit{ IEEE Communications Magazine}, vol. 53, no. 9, pp. 74-81, Sep. 2015.
\bibitem{kappa}
T. D. Burd and R. W. Brodersen, ``Processor Design for Portable Systems," \textit{Journal of VLSI signal processing systems for signal, image and video technology}, vol. 13, no. 2-3, pp. 203-221, Aug. 1996.
\bibitem{Tammer}
K. Tammer, ``The Application of Parametric Optimization and Imbedding to the Foundation and Realization of a Generalized Primal Decomposition Approach," \textit{Mathematical research}, vol. 35, pp. 376-386, 1987.
\bibitem{m_head_search}
A.R. Narayanan, A. Zela, T. Saikia, et al. ``Multi-headed Neural Ensemble Search," \textit{ArXiv abs/2107.04369}, 2021.
\bibitem{m_head_why}
S. Lee, S. Purushwalkam, M. Cogswell, D. Crandall, and D. Batra, ``Why m heads are better than one: Training a diverse ensemble of deep networks." \textit{ArXiv abs/1511.06314}, 2015.
\bibitem{ensemble_base_review}
M.A. Ganaie, M. Hu, A.K. Malik, M. Tanveer and P.N. Suganthan, ``Ensemble deep learning: A review," \textit{Engineering Applications of Artificial Intelligence}, vol. 115, no. 105151, 2022.
\bibitem{error_ambiguity_decomposition}
K. Anders and V. Jesper, ``Neural Network Ensembles, Cross Validation, and Active Learning,” \textit{MIT Press}, July 1994.


\end{thebibliography}
\end{document}